\documentstyle[12pt]{article}
\begin{document}
\title{Tests of isospin symmetry breaking at $\phi (1020)$ meson factories}
\author{H. Genz\\
Institut f\"{u}r Theoretische Teilchenphysik Universit\"{a}t Karlsruhe\\
D76128 Karlsruhe Germany\\ \\
\and
S. Tatur \\
N.Copernicus Astronomical Center,\\ Polish Academy of Sciences,\\
Bartycka 18, 00-716 Warsaw, Poland.}
\date{}
\maketitle
\begin{abstract}
\noindent In a model of isospin symmetry breaking we obtain the ($e^{-}
e^{+} \rightarrow \pi^{-} \pi^{+}$) amplitude $Q$ and the isospin $I=0$
and $I=1$ relative phase $\psi$ at the $\phi (1020)$ resonance in
aproximate agreement with experiment. The model predicts $\Gamma(\phi
\rightarrow \omega \pi^{0}) \approx 4 \cdot 10^{-4} \;\mbox{MeV}$.
We have also obtained $\Gamma (\phi \rightarrow \eta' \gamma)=5.2
\cdot 10^{-4} \;\mbox{MeV}$. Measuring this partial width would strongly
constrain $\eta$-$\eta'$ mixing. The branching ratios $BR$ of the isospin
violating decays $\rho^{+} \rightarrow \pi^{+} \eta$ and $\eta' \rightarrow
\rho^{\pm} \pi^{\mp}$ are predicted to be $BR(\rho^{+} \rightarrow \pi^{+}
\eta)=3 \cdot 10^{-5}$ and $BR(\eta' \rightarrow \rho^{\pm} \pi^{\mp})=4
\cdot 10^{-3}$, respectively, leading to $BR[\phi \rightarrow \rho^{\pm}
\pi^{\mp} \rightarrow (\pi^{\pm} \eta)\pi^{\mp} \rightarrow (\pi^{\pm}
\gamma \gamma)\pi^{\mp}]=10^{-6}$ and $BR[\phi \rightarrow \eta'
\gamma \rightarrow (\rho^{\pm} \pi^{\mp})\gamma]=2\cdot 10^{-6}$.
\end{abstract}
\newpage
\noindent {\bf 1. Introduction} \\

It is generally belived that electromagnetic interactions and the mass
differences of u and d quarks are the sources of the breaking of isospin
symmetry \cite{rev}. Both lead, among other things, to the mixing of the
isospin $I=0$ and $I=1$ members of the $SU(3)_{f}$ flavor nonets. The present
paper deals with isospin symmetry violating meson decays that proceed via
$\pi^{0}-\eta$, $\pi^{0}-\eta'$, $\omega-\rho^{0}$ and $\phi-\rho^{0}$
mixings. Well--known examples are the isospin--forbidden decays $\eta
\rightarrow 3\pi$ (the main decay channel of the $\eta$) and $\psi'\rightarrow
\psi \pi^{0}$ \cite{2}. We concentrate on the decays $\phi \rightarrow \pi^{+}
\pi^{-}$, $\phi \rightarrow \omega \pi^{0}$, $\rho^{\pm} \rightarrow \pi^{\pm}
\eta$ and $\eta' \rightarrow \rho^{\pm} \pi^{\mp}$ that can thoroughly
be investigated at $\phi$ meson factories. This is obvious for $\phi
\rightarrow \pi^{+} \pi^{-}$ and $\phi \rightarrow \omega \pi^{0}$. Since
the $\phi$ decays into $\rho \pi$ with a branching of $13\%$, $\rho$
mesons will be produced at $\phi$ meson factories with a rate that suffices
to detect and investigate $\rho^{\pm} \rightarrow \pi^{\pm} \eta$. The $\phi$
is furthermore expected to decay into $\eta' \gamma $ with a branching of
approximately $10^{-4}$. This will presumably be sufficient for detecting
and investigating $\eta' \rightarrow \rho^{\pm} \pi^{\mp}$.\\

\noindent {\bf 2. Input parameters} \\

We will use the matrix element
\begin{equation}
<\eta |H'|\pi^{0}>=-6000\;\mbox{MeV}^{2}.
\end{equation}
We quote two determinations of this value. It firstly follows from the recent
determination \cite{don}
\begin{equation}
\frac{m_{d}-m_{u}}{m_{s}}=1/29
\end{equation}
of the quark mass ratios, together with the formula
\begin{equation}
<\eta |H'|\pi^{0}>^{tadpole}=-\frac{m_{d}-m_{u}}{m_{s}}(m_{K}^{2}-
m_{\pi}^{2})x_{P}=-6350\;\mbox{MeV}^{2}
\end{equation}
This assumes the Zweig rule formula
\begin{equation}
_{PS}<s\bar{s} |H'|\pi^{0}>=0
\end{equation}
We assume \cite{pdg} $\theta_{P}=-20^{\circ}$ for the pseudoscalar meson
mixing angle and have defined
\begin{eqnarray}
|\eta>&=&x_{P}|\frac{u\bar{u}+d\bar{d}}{\sqrt{2}}>_{PS}+y_{P}|s
\bar{s}>_{PS}\\
|\eta'>&=&-y_{P}|\frac{u\bar{u}+d\bar{d}}{\sqrt{2}}>_{PS}+x_{P}|s
\bar{s}>_{PS}
\end{eqnarray}
such that $x_{P}=1/\sqrt{3}(cos\theta_{P}-\sqrt{2}sin\theta_{P})\approx
0.822$.
We furthermore have
\begin{equation}
<\eta |H'|\pi^{0}>=<\eta |H'| \pi^{0}>^{tadpole}+<\eta |
H'| \pi^{0}>^{el.}
\end{equation}
with \cite{ametl}
$<\eta |H'| \pi_{0}>^{el.}=520\;\mbox{MeV}^{2}$ leading to
\begin{equation}
<\eta |H'| \pi^{0}>=-5800\;\mbox{MeV}^{2}.
\end{equation}
Secondly, \cite{sca} has found
\begin{equation}
<\eta |H'| \pi^{0}>=(-5900 \pm 600)\;\mbox{MeV}^{2}.
\end{equation}
In view of the large errors, we neglect the electromagnetic contribution and
use the round number in eq.(1) i.e.
\begin{equation}
<\eta |H'| \pi^{0}> \approx <\eta |H'| \pi^{0}>^{tadpole} \approx
-6000\;\mbox{MeV}^{2}
\end{equation}
In the same way we assume\footnote{We also note that \cite{sca} has found
$<\eta' |H'|\pi^{0}>=(-5500\pm 500)\;\mbox{MeV}^{2}$ for $\theta_{P}=
-13^{\circ}$. Using eq.(9) together with these numbers one finds that $_{PS}<s
\bar{s} |H'| \pi^{0}>\approx 0$ also in the approach of that reference.}
\begin{eqnarray}
<\eta'|H'| \pi^{0}>\approx -y_{p}<\frac{u\bar{u}+d\bar{d}}{\sqrt{2}}|H'|
\pi_{0}>^{tadpole}=-\frac{y_{p}}{x_{p}}<\eta |H'| \pi^{0}> \nonumber \\
=-4200\;\mbox{MeV}^2.
\end{eqnarray}
We will also use \cite{13}
\begin{equation}
<\omega|H'|\rho^{0}>=(-4520\pm 600)\;\mbox{MeV}^2.
\end{equation}
For the $\omega$-$\phi$ mixing angle the Gell-Mann-Okubo value \cite{pdg}
$\theta_{V}=39^{\circ}$ will be used.

The couplings $f_{V}$ of the vector mesons $\rho$, $\omega$, $\phi$ to the
photon are defined such that
\begin{equation}
\Gamma (V\rightarrow e^{+}e^{-})=\frac{\pi m_{V}\alpha^{2}}{3}f_{V}^{-2},
\end{equation}
where $\alpha=1/137$ is the fine structure constant. Numerically
$f_{\rho}^{-2}=0.16$, $f_{\omega}^{-2}=0.014$, $f_{\phi}^{-2}=0.024$
in satisfactory agreement with the quark model relation $9:1:2$ for these
couplings. As also suggested by the quark model the $f_{V}$ are assumed to
be positive relative to each other.

One of the main ingredients of this work is the vector-vector-pseudoscalar
meson coupling constant $g$ defined by the effective Lagrangian
\begin{equation}
L_{I}=g/2\epsilon_{\alpha \beta \gamma \delta}\sum_{a,b,c=0}^{8}(\partial^
{\alpha}V^{\beta}_{a})V^{\gamma}_{b}(\partial^{\delta}P_{c})d_{abc}
\end{equation}
involving vector meson fields $V^{\alpha}_{a}$ with Lorentz-index $\alpha$,
$SU(3)_{f}$ index $a$ and the pseudoscalar fields $P_{a}$. The $d_{abc}$
are the well-known symmetric $SU(3)$ Clebsch-Gordan coefficients, $\epsilon_{
\alpha \beta \gamma \delta}$ is the four-dimensional antisymmetric $\epsilon$
tensor, and $\partial^{\alpha}$ is a differentiation symbol. The experimental
value of $\Gamma(\rho^{0} \rightarrow \pi^{0}\gamma)$ and the width formula
\begin{equation}
\Gamma(\rho^{0} \rightarrow \pi^{0} \gamma)=\frac{\alpha g^2}{94f_{\omega}^
{2}}(\frac{m_{\rho}^{2}-m_{\pi^{0}}^{2}}{m_{\rho}})^{3}
\end{equation}
imply $|g|=0.0164\;\mbox{MeV}^{-1}$. This value will be used below.

{}From the above, widths of the types $P\rightarrow\gamma\gamma$, $P\rightarrow
V\gamma$, $V\rightarrow P\gamma$, and $V\rightarrow PV$ can be computed
in overall agreement with experiment \cite{8,14,15}. Details of the results
depend on the assumed values of the meson mixing angles and $SU(3)_{f}$
symmetry breaking corrections and will not concern us here. As examples of
straightforward checks of our mixing assumptions, we calculate
\begin{eqnarray}
\Gamma(\rho^{0}\rightarrow\eta\gamma)&=&\frac{\alpha g^2}{96f_{\rho}^{2}}
(\frac{m_{\rho}^{2}-m_{\eta}^{2}}{m_{\rho}})^{3}(x_{P})^{2}\nonumber\\
&=&0.174(x_{P})^2\mbox{MeV}=0.119\;\mbox{MeV},\\
\Gamma(\eta'\rightarrow\rho^{0}\gamma)&=&\frac{\alpha g^2}{216f_{\rho}^{2}}
(\frac{m_{\eta'}^{2}-m_{\rho}^{2}}{m_{\eta'}})^{3}(y_{P})^{2}\nonumber\\
&=&0.058(y_{P})^{2}\;\mbox{MeV}=0.019\;\mbox{MeV},\\
\Gamma(\phi\rightarrow\rho\pi)&=&3\Gamma(\phi\rightarrow\rho^{0}\pi^{0})=
\frac{g^2}{4\pi}(q(\phi\rightarrow\rho^{0}\pi^{0}))^{3}(x_{V})^{2}\nonumber\\
&=&136(x_{V})^2\;\mbox{MeV}=0.55\;\mbox{MeV}
\end{eqnarray}
and
\begin{equation}
\Gamma(\phi\rightarrow\eta\gamma)=\frac{\alpha g^2}{48f_{\phi}^{2}}(\frac
{m_{\phi}^{2}-m_{\eta}^{2}}{m_{\phi}})^{3}(y_{P})^{2}\nonumber\\
=0.38(y_{P})^{2}\;\mbox{MeV}=0.122\;\mbox{MeV}.
\end{equation}
Comparison to the experimental values $\Gamma(\rho^{0}\rightarrow\eta\gamma)=
0.058\;\mbox{MeV}$, $\Gamma(\eta'\rightarrow\rho^{0}\gamma)=0.059\;\mbox{MeV}
$, $\Gamma(\phi\rightarrow\rho^{0}\pi^{0})=0.57\;\mbox{MeV}$, and $\Gamma
(\phi\rightarrow\eta\gamma)=0.057\;\mbox{MeV}$ indicates the quality of
agreement to be expected {\em without corrections}. In particular, the poor
agreement in the only case where $SU(3)_{f}$ is applied to $s\bar{s}$ states
(i.e.$\phi\rightarrow\eta\gamma)$ invites an $SU(3)_{f}$ symmetry breaking
correction \cite{8}. Overall fits \cite{8,14,15} of course also improve the
apparent agreement. Theoretical and experimental $\Gamma(\phi\rightarrow
\rho^{0}\pi^{0})$ compare surprisingly well, lending support to our choise of
$\theta_{V}$.

The width $\Gamma(\phi\rightarrow\eta'\gamma)$ can be written in terms of
$\Gamma(\phi\rightarrow\eta\gamma)$ as
\begin{equation}
\Gamma(\phi\rightarrow\eta'\gamma)=(\frac{x_{P}}{y_{P}})^{2}(\frac{m_{\phi}^
{2}-m_{\eta'}^{2}}{m_{\phi}^{2}-m_{\eta}^{2}})^{3}\Gamma(\phi\rightarrow\eta
\gamma)=5.2\cdot 10^{-4}\;\mbox{MeV}.
\end{equation}
The numerical value follows from assuming that $\eta$ and $\eta'$ do not mix
with mesons outside their $SU(3)_{f}$ nonet.\\

\noindent {\bf 3. The decay $\phi\rightarrow\pi^{+}\pi^{-}$} \\

The observed decays of the $\phi (1020)$ into purly hadronic final states are
$\phi\rightarrow K\bar{K}$ (Fraction: $0.83$), $\rho\pi$ ($0.13$), $\pi^{+}
\pi^{-}\pi^{0}$ ($0.024$), and $\pi^{+}\pi^{-}$ ($8\cdot 10^{-5}$).
This pattern can be understood if the physical $\phi$ is an almost ideally
mixed $s\bar{s}$ vector meson (i.e. $I^{G}=0^{-}$) that decays into $\pi^{+}
\pi^{-}$ electromagnetically (Fig.1c) as well as hadronically via an isospin
symmetry violating mixing with $\rho^{0}$ (Fig.1b). More specifically, we
assume the resonant part $A_{\phi}$ of the invariant $<\gamma|\pi^{+}\pi^{-}>$
amplitude (Fig.1a) around $\sqrt{s}=1020\;\mbox{MeV}$ to be given by Fig.1b
and 1c. The total amplitude
\begin{equation}
A(s)=F_{\pi}(s)+A_{\phi}(s)
\end{equation}
also contains a non-resonant smooth interpolation $F_{\pi}(s)$ of the total
formfactor $A$ of the $\pi$ over the $\phi$ resonance region. As an
illustrative examle we will saturate $F_{\pi}(s)$ by the Breit-Wigner
propagator of the $\rho$ in the normalization $F_{\pi}(m_{\rho}^{2})=1/
i\Gamma_{\rho}m_{\rho}$ such that
\begin{equation}
F_{\pi}(s)=\frac{1}{s-m_{\rho}^{2}+i\Gamma_{\rho}m_{\rho}}.
\end{equation}
Since $A_{\phi}(m_{\rho}^{2})\approx 0$, we also have $A(m_{\rho}^{2})\approx
F_{\pi}(m_{\rho}^{2})=1/i\Gamma_{\rho}m_{\rho}$. From Figs. 1b and 1c we read
off
\begin{equation}
A(s)=F_{\pi}(m_{\phi}^{2})+\frac{f_{\rho}}{f_{\phi}}\frac{m_{\phi}^{2}}
{m_{\rho}^{2}}\frac{1}{s-m_{\phi}^{2}+i\Gamma_{\phi}m_{\phi}}[\frac{x_{V}
<\omega|H'|\rho^{0}>}{s-m_{\rho}^{2}+i\Gamma_{\rho}m_{\rho}}+\frac{\alpha\pi}
{f_{\phi}f_{\rho}}F_{\pi}(m_{\phi}^{2}]
\end{equation}
for $s\approx m_{\phi}^{2}$. In analogy to the pseudoscalar case we have
defined $x_{V}=1/\sqrt{3}(cos\theta_{V}-\sqrt{2}sin\theta_{V})$. \\

A few comments are in order. Since the hadronic isospin symmetry breaking
Hamiltonian is proportional to ($u\bar{u}-d\bar{d}$), the $\phi$ couples
hadronically to the $\rho$ via $\phi$-$\omega$ mixing only. This yields the
factor $x_{V}$. The contribution of $\omega\rightarrow\gamma\rightarrow\rho
^{0}$ is contained in $<\omega|H'|\rho^{0}>$. The contribution of the $\rho^
{0}$ at $\sqrt{s}\approx m_{\phi}$ is then parametrized by the Breit-Wigner
propagator. This is entirely correct for the hadronic $\omega$-$\rho^{0}$
transition. For the photonic transition $\omega\rightarrow\gamma\rightarrow
\rho^{0}$ it may be argued that the coupling is to $F_{\pi}(m_{\phi}^{2})$
rather than to the Breit-Wigner $\rho^{0}$. It is however easy to see that
the photonic contribution to $<\omega|H'|\rho^{0}>$ is small. Namely,
 the contribution of $\omega\rightarrow\gamma\rightarrow
\pi^{-}\pi^{+}$ to $\omega\rightarrow\pi^{-}\pi^{+}$ can be read off an
obvious modification of eq.(23). If there were no other contribution, $\Gamma
(\omega\rightarrow\pi^{-}\pi^{+})=0.005\;\mbox{MeV}$ (whereas the experimental
width is $0.19\;\mbox{MeV}$). We follow \cite{5} in factorizing the $<\gamma
|\pi^{+}\pi^{-}>$ amplitude in the neighbourhood of the $\phi (1020)$ such
that the $s$ dependent factor containing the $\phi$ resonance reads
\begin{equation}
1+Q\frac{e^{i\psi}m_{\phi}\Gamma_{\phi}}{s-m_{\phi}^{2}+im_{\phi}\Gamma_
{\phi}}\nonumber\\
=1+\frac{f_{\rho}}{f_{\phi}}\frac{m_{\phi}^{2}}{m_{\rho}^{2}}\frac{Y}{s-
m_{\phi}^{2}+im_{\phi}\Gamma_{\phi}},
\end{equation}
where Y is given by
\begin{equation}
Y=\frac{\alpha\pi}{f_{\rho}f_{\phi}}m_{\rho}^{2}+|\frac{F_{\pi}(m_{\rho}^
{2})}{F_{\pi}(m_{\phi}^{2})}|\frac{m_{\rho}\Gamma_{\rho}x_{V}<\omega|H'|\rho^
{0}>}{m_{\phi}^{2}-m_{\rho}^{2}+i\Gamma_{\rho}m_{\rho}}e^{iR},
\end{equation}
with $R$ the phase of $F_{\pi}(m_{\phi}^{2})$ at the $\phi$ resonance
\begin{equation}
F_{\pi}(m_{\phi}^{2})=|F_{\pi}(m_{\phi}^{2})|e^{-iR}.
\end{equation}
Assuming for $m_{\phi}$ and $\Gamma_{\phi}$ the values of \cite{pdg} $m_
{\phi}=1019.4\;\mbox{MeV}$ and $\Gamma_{\phi}=4.43\;\mbox{MeV}$,
respectively, the observables to be determined experimentally are $\psi$ and
\begin{equation}
Q=[\frac{36B(\phi\rightarrow\pi^{-}\pi^{+})B(\phi\rightarrow e^{-}e^{+})}
{\alpha^{2}(1-4m_{\pi}^{2}/m_{\phi}^{2})^{3/2}|F_{\pi}|^{2}}]^{1/2}.
\end{equation}
Ref.\cite{5} finds
\begin{equation}
\psi=(20\pm 13)^{\circ}
\end{equation}
together with
\begin{equation}
Q=0.07\pm0.02.
\end{equation}
Using, as has also been obtained in\cite{5}, $|F_{\pi}|^{2}=2.9\pm 0.2$ and
$B(\phi\rightarrow e^{-}e^{+})=3\cdot 10^{-4}$ this yields
\begin{equation}
\Gamma(\phi\rightarrow\pi^{+}\pi^{-})=2.8\cdot 10^{-4}\;\mbox{MeV}.
\end{equation}
The value $\Gamma(\phi\rightarrow\pi^{+}\pi^{-})=(3.5\begin{array}{c}+2.2\\
-1.8\end{array})\cdot 10^{-4}\;\mbox{MeV}$ \cite{pdg} combines this result with
the much
higher $\Gamma(\phi\rightarrow\pi^{+}\pi^{-})=8\cdot 10^{-4}\;\mbox{MeV}$ of
ref.\cite{6}

Using $R$ as a parameter, our results are presented in Table 1. Approximate
agreement with experiment is obtained for $R$ between $-170^{\circ}$ and
$15^{\circ}$. A more meaningful test will hopefully be provided by DA$\Phi$NE.
\\

\noindent {\bf 4. The decay $\phi\rightarrow\omega\pi^{0}$} \\

The amplitude $<\phi|\omega\pi^{0}>$ is determined by the contributions of
Fig.2. We emphasize that, other than in Fig. 1c, we have in Fig.2c {\em only
to take the contribution of the $\rho^{0}(770)$} into account, i.e. {\em not}
the full $F_{\pi}(M_{\phi}^{2})$. Namely interpreting $F_{\pi}(s)$ in the
neighbourhood of the $\phi (1020)$ as the {\em sum}  of the contributions of
the ($I=1$) vector mesons $\rho (770)$, $\rho (1450)$, and $\rho(1700)$, the
Meson Full Listings of Ref.\cite{pdg} suggest that {\em only the $\rho(770)$
considerably couples to $\omega\pi^{0}$}. Thus we may write for the amplitude
(taking the $\rho^{0} \omega\pi^{0}$ coupling from Sect.2 rather than
directly from the Gell-Mann-Sharp-Wagner calculation \cite{15,17} of
$\omega\rightarrow 3\pi )$
\begin{equation}
A=\sum_{j=a}^{e} A^{(j)}
\end{equation}
with
\begin{eqnarray}
A^{(a)}&=&\frac{gx_{V}<\omega|H'|\rho^{0}>}{m_{\omega}^{2}-m_{\rho}^{2}+im_{
\rho}\Gamma_{\rho}},\\
A^{(b)}&=&\frac{gx_{V}<\omega|H'|\rho^{0}>}{m_{\phi}^{2}-m_{\rho}^{2}+
i\Gamma_{\rho}m_{\rho}},\\
A^{(c)}&=&\frac{\alpha g\pi m_{\rho}^{2}}{f_{\phi}f_{\rho}}\frac{1}{m_{\phi}^
{2}-m_{\rho}^{2}+i\Gamma_{\rho}m_{\rho}},\\
A^{(d)}&=&x_{V}g[x_{P}\frac{<\eta|H'|\pi^{0}>}{m_{\pi^{0}}^{2}-m_{\eta}^{2}}-
y_{P}\frac{<\eta'|H'|\pi^{0}>}{m_{\pi^{0}}^{2}-m_{\eta'}^{2}}]
\end{eqnarray}
and
\begin{equation}
A^{(e)}=\sqrt{2}x_{V}g[y_{P}\frac{<\eta|H'|\pi^{0}>}{m_{\pi^{0}}^{2}-m_{\eta}
^{2}}+x_{P}\frac{<\eta'|H'|\pi^{0}>}{m_{\pi^{0}}^{2}-m_{\eta'}^{2}}]
\end{equation}
The width is
\begin{equation}
\Gamma(\phi\rightarrow\omega\pi^{0})=|A|^2\frac{(q(\phi\rightarrow\omega
\pi^{0}))^3}{12\pi}
\end{equation}
The input parameters we use yield $\Gamma(\phi\rightarrow\omega\pi^{0})=4\cdot
10^{-4}\;\mbox{MeV}$.
\\

\noindent {\bf 5. The decays $\rho^{\pm}\rightarrow\pi^{\pm}\eta$
and $\eta '\rightarrow \rho^{\pm}\pi^{\mp}$} \\

Finally we consider the decays $\rho^{\pm}\rightarrow\pi^{\pm}\eta$ and
$\eta'\rightarrow\rho^{\pm}\pi^{\mp}$. They obviously violate $G$ parity and
thus isospin symmetry ( since charge conjugation symmetry holds for these
strong and/or electromagnetic decays). The strong isospin symmetry conserving
matrix element leading to the proposed decays via $\pi^{0}\eta$ and
$\pi^{0}\eta'$ mixing, is in both cases $<\rho^{\pm}\pi^{\mp}\pi^{0}>$.

With $q$ the decay momenta of the $P$-wave decays we thus obtain
\begin{equation}
\Gamma(\rho^{+}\rightarrow\pi^{+}\eta)=[\frac{q(\rho\rightarrow\pi\eta)}
{q(\rho\rightarrow\pi\pi)}]^3\cdot [\frac{<\eta|H'|\pi^{0}>}{m_{\pi}^{2}-m_
{\eta}^{2}}]^2\cdot \Gamma(\rho^{+}\rightarrow\pi^{+}\pi^{0})=4\cdot 10^{-3}\;
\mbox{MeV}
\end{equation}
and, taking both final charge states together
\begin{eqnarray}
\Gamma(\eta'\rightarrow\rho^{\pm}\pi^{\mp})=2\cdot3\cdot[\frac{q(\eta'
\rightarrow\rho\pi)}{q(\rho\rightarrow\pi\pi)}]^3\cdot[\frac{<\eta'|H'|\pi^
{0}>}{m_{\pi}^{2}-m_{\eta'}^{2}}]^2\cdot\Gamma(\rho^{+}\rightarrow\pi^{+}
\pi^{0}) \nonumber \\
=7\cdot10^{-4}\;\mbox{MeV}.
\end{eqnarray}
The factor $3$ in the second formula is due to the summation (rather than
averaging) over the spin orientations of the $\rho$.
The results for the branching ratios $BR$  are $BR(\rho^{+}\rightarrow\pi^{+}
\eta)=3\cdot 10^{-5}$ and $BR(\eta'\rightarrow\rho^{\pm}\pi^{\mp})=4\cdot
10^{-3}$, leading to branchings  $BR[\phi \rightarrow \rho^{\pm}
\pi^{\mp} \rightarrow (\pi^{\pm} \eta)\pi^{\mp} \rightarrow (\pi^{\pm}
\gamma \gamma)\pi^{\mp}]=10^{-6}$ and $BR[\phi \rightarrow \eta'
\gamma \rightarrow (\rho^{\pm} \pi^{\mp})\gamma]=2\cdot 10^{-6}$. Thus
meaningfull conclusions on these decays can be obtained at $\phi$ meson
factories \cite{kluge}.
\\

\noindent {\bf 6. Conclusions} \\

In conclusion we note that our understanding of the processes considered here
and those connected to them by $SU(3)_{f}$ symmetry, vector meson dominance,
and isospin symmetry breaking will be strongly enhanced by accurate $e^{-}
e^{+}$ annihilation experiments in the $\phi(1020)$ resonance region. As to
orders of magnitude, mainstream low energy phenomenology can accomodate the
observed $\psi=(-20\pm 13)^{\circ}$ and $Q=0.07\pm 0.02$ within the large
experimental errors. Our input data favor a $\psi$ that is negative and nearer
to $0^{\circ}$ and/or a larger $Q$ (implying a larger $\Gamma(\phi\rightarrow
\pi^{-}\pi^{+}))$. The width $\Gamma(\phi\rightarrow\omega\pi^{0})$ is
predicted to be approximately $4\cdot 10^{-4}\;\mbox{MeV}$.
Our predictions concerning $\rho^{\pm}\rightarrow\pi^{\pm}\eta$
and $\eta '\rightarrow \rho^{\pm}\pi^{\mp}$ are stated at the end of
the previous section.
\vspace{3mm} \\

{\bf Acknowledgments}\\

In an early stage of this work, Dr. J. Iqbal has approved of the methods of
isospin symmetry breaking we use. Valuable hints by him and Dr.M.~Scadron
at the literature are gratefully acknowledged. We acknowledge fruitful
discussions with Dr.R.~Decker. This work has been started
during a sabbatical stay of one of the authors (H.G.) at TRIUMF. He would
like  to thank Dr.H.~Fearing and the Theory Group for their kind hospitality.
The visit was made possible by a grant of the Stiftung Volkswagenwerk, which
is gratefully acknowledged.

\newpage
\noindent {\bf Figure Captions}\\
Fig.1. Diagrams used to compute $\psi$ and $Q$ of $e^{-}e^{+}\rightarrow\pi^{
-}\pi^{+}$ near the $\phi(1020)$ resonance.\\
Fig.2 Diagrams used to compute the width $\Gamma(\phi\rightarrow\omega\pi^{0})
$\\

\noindent Table 1. Using $R$ of eq.(26) as parameter, the predicted values
of $\psi$ and $Q$ (eq.(27)) are listed and compared to experiment \cite{5}.\\

\begin{tabular}{lccc} \hline\hline
Condition                & $R$ (degree) & $\psi$ (degree) & $Q$ \\ \hline
Experiment               &              & -20 $\pm$ 13     & 0.07$\pm$0.02\\
Destructive interference & -165.7       & 0               & 0.089\\
                         & -150         & -6.4            & 0.091\\
                         & -120         & -15             & 0.10\\
Phase $\psi$ minimal     & -93          & -17             & 0.12\\
Consructive interference & 14.3         & 0               & 0.16\\
Phase $\psi$ maximal     & 122          & 17              & 0.12\\ \hline
\hline
\end{tabular}
\end{document}